\documentclass[prl,aps,floatfix,nofootinbib,preprint]{revtex4} 
\usepackage{graphicx}
\usepackage{amssymb} 
\usepackage{amsmath} 
\usepackage{amsfonts} 
\usepackage{amsbsy}
\usepackage{color}

\def\be{\begin{equation}} 
\def\ee{\end{equation}}

\begin{document} 

\title{Exit-Channel Suppression in Statistical Reaction Theory
}
\author{G.F.~Bertsch$^{1}$ and T.~Kawano$^{2}$}
 
\affiliation{$^{1}$Department of Physics and Institute of Nuclear Theory, 
University of Washington, Seattle, Washington 98915, USA\\ 
$^{2}$Theoretical Division, Los Alamos National Laboratory, 
Los Alamos, New Mexico 87545, USA}
 
\begin{abstract}

Statistical reaction theories such as Hauser-Feshbach  
assume that branching ratios follow Bohr's compound nucleus hypothesis
by factorizing into independent probabilities for 
different channels. Corrections
to the factorization hypothesis are known in both nuclear theory and
quantum transport theory, particularly an enhanced memory of the
entrance channel.  
We apply the Gaussian orthogonal ensemble to study a complementary 
suppression of exit channel branching ratios.  The combined effect of the
width fluctuation and the limitation on the transmission coefficient 
can provide a lower bound on the number of exit channels.
The bound is demonstrated 
for the branching ratio in neutron-induced reactions on a 
$^{235}$U target. \\ 
\end{abstract}
\maketitle

{\it Introduction.} Statistical approximations are extremely useful in nuclear physics,
particularly in reaction theory.  Examples are the  
Hauser-Feshbach and Weisskopf-Ewing formulas for reaction cross sections
\cite{es06,we40,ha52}.  
The underlying assumption of both is the factorizability of the cross section
$\sigma_{ab}$ from one channel to another as $\sigma_{ab} \sim \Gamma_a \Gamma_b$, 
where $\Gamma_i$ is the average decay width through a channel.  The factorization
follows from  Bohr's compound nucleus hypothesis \cite{bo36},
that the decay of a heavy nucleus
has no memory of how it was formed.  However, factorization is only 
justified when the reaction takes place through discrete resonances
and there are many channels contributing to each decay mode. Otherwise,
the  fluctuations in the widths of the resonance gives rise to the
well-known ``width fluctuation correction" (WFC) to the statistical models 
\cite{mo75}, most prominently as the ``elastic enhancement factor."  There
is now an extensive literature on the subject cited in Refs. \cite{mi10} 
and \cite{ka15}.  
Similar effects in electron propagation through mesoscopic conductors
are known as the  ``weak localization correction" 
and the ``dephasing" effect \cite[Sect. IV.C and IV.E]{al00}.

While the nuclear correction is best known as an entrance channel effect,
it can also be present in exit channels if the reaction branching ratios
highly favor a decay mode with large fluctuations \cite{mo76}. 
In this work we show that such situations can lead to effects large enough
to provide bounds on the number of channels in the decay mode, even though
the measurements are on averaged quantities and not on their fluctuations.
This study was motivated in part by the quest for a theory of
fission dynamics based on nucleon-nucleon interactions.  That requires
an understanding not only of the distribution of the fission
channels but their coupling matrix elements to the other states.

{\it The GOE statistical model.}
The factorization hypothesis and other statistical aspects of reaction
theory  can be tested theoretically by models that consider ensembles of 
Hamiltonians that  mix the constituent 
configurations. The Gaussian orthogonal ensemble (GOE) has been especially
successful in this regard \cite{ka15,mi10}.  The reaction theory is
expressed in the matrix equations
\be
K = \pi { \gamma}^T {1\over E - H} { \gamma}
\ee 
\be
S = {1 -i  K \over 1 +i  K} 
\ee
giving for the non-elastic cross sections \cite{statistical}
\be
\sigma_{nf} = {\pi\over k_n^2} \sum_{c \in f} |S_{n c}|^2 \ .
\ee
In Eq. (1) $K$ is a matrix of dimension $N^{ch}\times N^{ch}$, where $N^{ch}$ is 
the number of reaction channels in the model. $H$ is the $N_\mu\times
N_\mu$ Hamiltonian matrix for the $N_\mu$ internal states 
in the model. The internal states are connected to the channels by the
$N^{ch} \times N_\mu$ reduced-width matrix $ \gamma$.
Eq. (2) relates the $K$-matrix
to the familiar $S$-matrix of scattering theory. There is an additional
overall phase factor in Eq. (2) which plays no role in the reaction
cross sections.  In Eq.~(3) 
 $n$ is the entrance channel, $f$ is a set of exit
channels that are grouped together in an experimental cross section, and $c$ are
the individual channels. The cross section depends explicitly on
the entrance channel energy $E_n$ via { the neutron wave-number} $k_n = \sqrt{2 E_n M_n}$, with $M_n$ 
the reduced mass.

In the GOE statistical model, the Hamiltonian $H$ is sampled from the
distribution {\cite{me04}}
\be
H_{\mu,\mu'} = H_{\mu',\mu} = v_{\mu,\mu'} \left( 1 +
\delta_{\mu,\mu'}\right)^{1/2} \ ,
\ee  
where $\mu \ge \mu'$ and $v_{\mu,\mu'}$ is a Gaussian-distributed random variable.  The ensemble is
completely specified by $N_\mu$ and the r.m.s.~Hamiltonian matrix element 
$\langle v^2\rangle^{1/2}$. 
Here we shall
characterize the GOE ensemble by $D$, the average level spacing in the middle
of the distribution. The spacing is related to the matrix elements by $D = \pi 
\langle v^2\rangle^{1/2} N_{\mu}^{-1/2} $.
%
The
$  \gamma$ matrix associated with a GOE Hamiltonian can be assumed to 
have a diagonal structure of the
form
$ \gamma|_{\mu,c} = \gamma_c \delta_{\mu,c}$.  In this work, we also make the
simplifying assumption that
the  $\gamma_c$ are equal for all 
channels within a given decay mode $f$. 
When the matrix is transformed
to the basis diagonalizing the GOE Hamiltonian, the amplitudes will be 
distributed over eigenstates according the Porter-Thomas distribution\,
\cite{PT} 
with a number of degrees of freedom equal to the number of channels $N_f$.  
It will be 
convenient to define an effective $K$-matrix decay rate for the different
modes $\Gamma^K_f$ as
\be
\Gamma^K_f =  {2 \pi \over  N_\mu}{ \sum_{c \in f} } \gamma_c^2 \ .
\ee
If all the $\Gamma_f^K$ are small compared to $D$, 
the average $S$-matrix decay widths satisfy 
\be
\Gamma_f \approx \Gamma_f^K.
\label{Gamma_f}
\ee
Note also that $N^{ch}_n = 1$ for the entrance channel; 
its reduced width controls the total reaction cross section\,\cite{Js}.

The calculations reported below were carried out with codes that constructed the
GOE distribution by Monte Carlo sampling of $H$ and applying {Eqs}.~(1-3).  The
codes and input data are provided in the Supplementary Material \cite{sup}.

{\it Application to the branching ratio
$^{235}$U(n,f)/$^{235}$U(n,$\gamma$).}
Here we show by a physical example that cross-section branching ratios that
heavily favor some particular exit channel can be severely suppressed.
The behavior follows from
the GOE statistical model as formulated in the last section and is
thus universal.  Our example is the neutron-induced reactions on $^{235}$U.  
For neutron energies below
$\sim 10$ keV the predominant reactions are in the $s$-wave leading to
capture by gamma emission or fission.
An important quantity is $\alpha^{-1}$, the ratio of the
fission cross section $\sigma_F$ 
to capture cross section $\sigma_{cap}$, $\alpha^{-1} =
\sigma_F/\sigma_{cap}$.  It varies in the range $\alpha^{-1} \sim $ 2-3 
in the 1 - 15 keV energy range\cite{average}. 
The capture widths can assumed
to be constant over this energy interval, since the interval is very small compared
to the excitation energy ($\sim 6.5$ MeV) and capture takes place through
many channels. The same is true of the level density, since the effective
temperature is of the order of 0.5 MeV. Table I gives approximate values of measured
$D$ and $\Gamma_{cap}$ which will be used to determine the parameters of the 
$K$-matrix \cite{D}
\begin{table}[htb] 
\begin{center} 
\caption{ Experimental observables for neutron-induced reactions on
$^{235}$U.  The cross section data is at a neutron bombarding energy
$E_n = 10 $ keV.  The last two entry are the ratio of cross sections,
show the range of the ratios as well as the value at 10 keV.}
\label{experiment}
\begin{tabular}{|ccc|} 
\hline 
  Observable  & Value & Source\\
\hline
 $D$     &  $0.45\pm0.05$  eV  &  \cite{ripl,mug}  \\
$\Gamma_{cap}$ &   $38\pm 3$  meV  &  \cite{ripl} \\
$ \sigma_{cap}$  &  $1.05\pm 0.07$ {{b}}  & \cite{endf,corvi} \\
$ \sigma_F $   &    $2.96\pm 0.06$  {b}  & \cite{endf,corvi} \\
$\alpha^{-1}(10)$ &  $2.8\pm 0.2$ & \cite{corvi,footnote} \\
$\alpha^{-1}(1-15)$ &  $2$-$3.1$  & \cite{endf} \\

\hline 
\end{tabular} 
\end{center} 
\end{table} 
We will examine the cross section at $E_n = 10 $ keV; the experimental
values averaged over a 1 keV are also given in Table I.  

The $K$-matrix reduced-width parameters are determined as follows.
The
capture width is small compared to $D$ and many channels contribute so we 
can safely apply Eq.~(\ref{Gamma_f}); the equivalent $\Gamma_{cap}$ is shown in
Table II. The coupling to the
entrance channel depends on $E_n$ 
and is usually parameterized by the strength function
$S_0$ as
\be
{\langle \Gamma_n \rangle \over D}  = S_0  E^{1/2}.
\ee
From total cross sections one finds $S_0 \approx 1\pm0.1\times10^{-4}$ eV$^{-1/2}$
\cite{mo78,mug},\cite[Fig. 47]{ko03} and we use that value to determine 
the entry in Table II.  We note that this value is consistent with the
coupled-channel analysis of Ref. \cite{so05}.  For the fission reduced
width, we first make the factorization (Hauser-Feshbach) approximation and  
assume that the nominal decay widths scale with the cross sections, i.e.
$\Gamma^K_F/\Gamma^K_{cap}= \sigma_{F}/\sigma_{cap}$.
\begin{table}[htb] 
\caption{ $K$-matrix parameters (eV) describing observed  
cross sections 
at $E_n = 10 $ keV in Hauser-Feshbach theory 
and assuming  $\Gamma << D$.  We have also include the parameters from
the ENDF/B-VII.1 evaluation.
}
\label{K-parameters2}
\begin{tabular}{|c|ccc|} 
\hline 
     &   $\Gamma^K_{cap}$    & $\Gamma^K_F$   & ${\Gamma^K_n}$   \\
\hline
this work &   $  0.038\pm 3$    & $ 0.105\pm0.01$  &  $0.010\pm0.001 $\\ 
 ENDF   & $  0.039 $    & $ 0.289$  &  $0.0097 $\\ 
\hline 
\end{tabular} 
\end{table} 
 
The number of channels and states in the $K$-matrix are still to be 
specified.  As presented, the model is independent of the number
of states as long as that number is large.  We shall
take $N_\mu=50-100$.  The capture channels do not show large fluctuations
and one can therefore assume that $N^{ch}_{cap}>>1$; we take
$N^{ch}_{cap} = 10$ in our modeling.
The number of fission
channels is not well known { \cite{si16}} and we consider two possibilities:  model A with
one fission channel and model B with five fission channels.  

With all parameters now specified in the GOE $K$-matrix, we can 
compute the average cross sections and branching ratios.  These are
shown in Table III.  
\begin{table}[htb] 
\caption{
Average reaction cross sections at $E_n = 10$~keV, comparing models A
and B with experiment.  The uncertainties on the calculated values are
the r.m.s. sample-to-sample fluctuations associated with the random matrix
ensemble of the internal states, taking a 1 keV averaging interval.
We have also included in the table the impact on the elastic scattering 
$S$-matrix.
}
\label{results}
\begin{center} 
\begin{tabular}{|c|ccccc|} 
\hline 
 & $N^{ch}_F$  &  $\sigma_F$ (b)  &  $\sigma_{cap}$ (b)  & $\alpha^{-1}$ & $|S_{nn}|^2$  \\ 
\hline
 Exp.  &  & $2.96\pm 0.21$   & $1.06\pm0.07$  & 2.8 &    \\
  A    &  1 & $1.66\pm0.05$ & $1.39\pm0.05$ & $1.20\pm0.07$ & 0.954 \\
  B    &  5 & $2.28\pm0.10$ & $1.00\pm0.06$ & $2.3\pm0.11$ & 0.950 \\
\hline 
\end{tabular} 
\end{center} 
\end{table} 
\begin{table}[htb] 
\caption{
Average fission widths $\Gamma^K_F$ required to reproduce the observed cross-section
ratio $\alpha^{-1} = 2.8\pm0.2$ in various models.  HW:  Hauser-Feshbach;
HW/WFC;  Hauser-Feshbach with width fluctuation correction from Eq.
(\ref{WFC}); KtoS:  Eqs. (1-3).  The uncertainty of the observed
$\alpha^{-1}$ is propagated through the models to give the uncertainty
bars in the table.  Units are eV.
}
\label{models}
\begin{center} 
\begin{tabular}{|ccc|ccc|} 
\hline 
 Model & $N^{ch}_{cap}$  &  $N^{ch}_F$ &    HW  &  HW/WFC  & KtoS  \\ 
\hline
A & 10 & 1 & $0.107\pm0.008$ & $0.46\pm0.06$ &  none \\
B & 10 & 5 & $0.107\pm0.008$ & $0.13\pm0.01$ &  $0.136\pm0.01$\\
\hline 
\end{tabular} 
\end{center} 
\end{table} 
In model A, one sees an enhancement of the capture cross section and
a corresponding suppression of the fission cross section.  Clearly
factorization is violated.

Let us see if we can reproduce
the experimental branching ratio simply by increasing the fission width,
but keeping only a single channel.  Taking the width 
as a free parameter, we obtain the branching ratios shown as the
solid line in Fig. 1. Also, we show in Table IV the values of $\Gamma^K_F$ 
required to fit the observed branching ratio.  Model A saturates at
\begin{figure}[tb]
\includegraphics[width=8 cm]{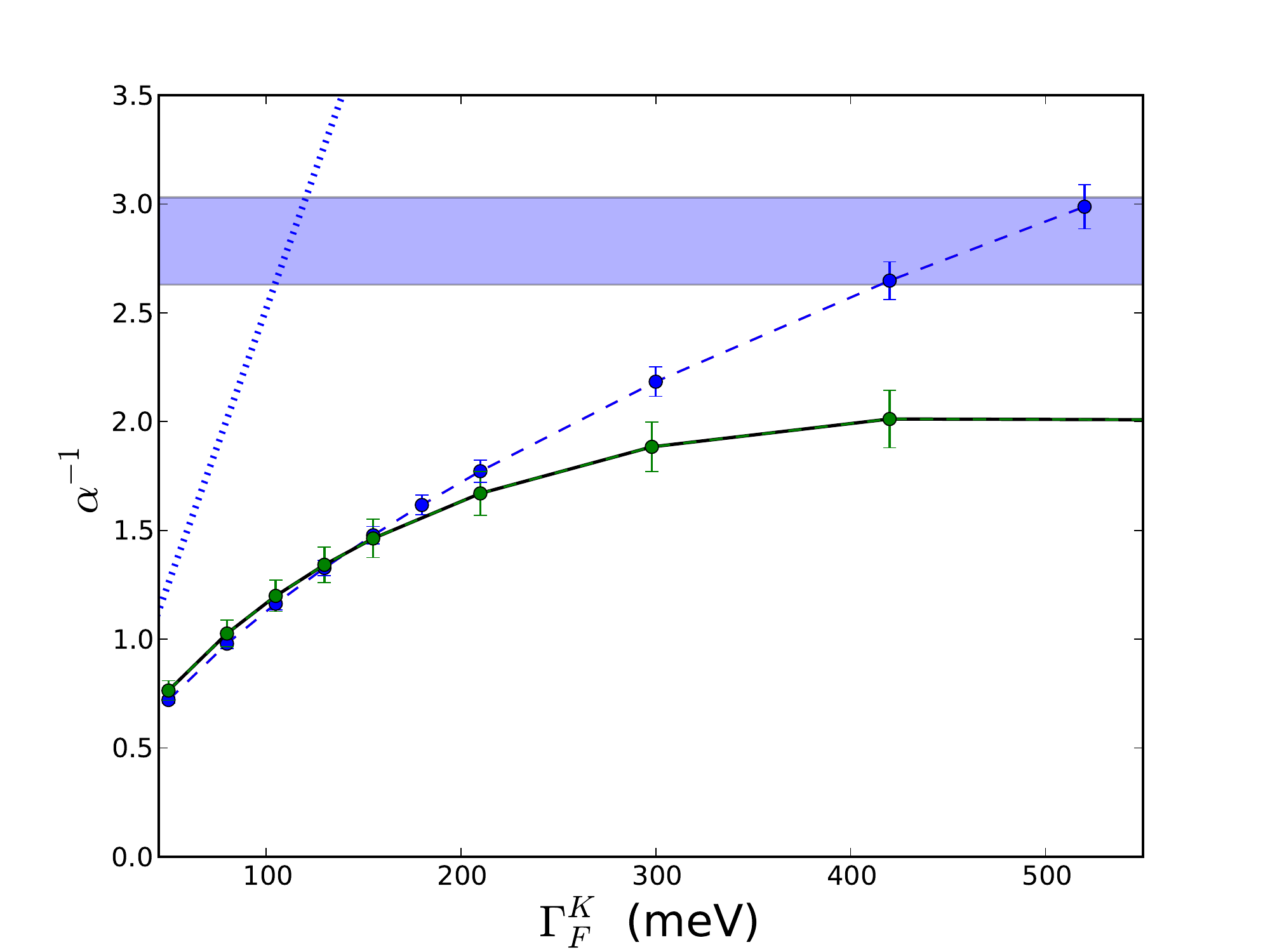}
\caption{Cross section ratio 
$\alpha^{-1}= \langle\sigma_F\rangle/\langle\sigma_{cap}\rangle$ as a function of
average fission width $\Gamma_F^K$ assuming a single 
fission channel.  Solid line:  Eq. (1-3). 
Dotted line:  Hauser-Feshbach approximation, i.e. $\alpha^{-1} = \Gamma_F^K/
\Gamma_{cap}^K$.  Dashed line:  Hauser-Feshbach including the WFC
correction, Eq. (\ref{WFC}). Blue band:  experimental range, taking uncertainty from
Table I.  
Widths are
the statistical errors associated with the 1 keV cross-section averaging 
interval. 
}
\label{ratio}
\end{figure}
$\alpha^{-1} \sim 2.0$;  there is no
reasonable value of $\Gamma^K_F$ that can reproduce experiment.  Thus, we can exclude fission 
models having only a single channel, based solely on average cross section data.
Of course, the fluctuation in cross-section ratios also carries information
on the number of channels and is the basis of previous estimates that
the effect channel count is of the order of a few. Finally, one can see from the
second line of Table IV that model B can fit the data taking the decay with
close to the Hauser-Feshbach value.

{\it Discussion.}
The exit channel suppression comes about by two mechanisms that can be
understood as follows.  The part coming from Porter-Thomas fluctuations
	can be analyzed at the level of the $K$-matrix:  assuming isolated
resonances, the branching ratio can be calculated as in Ref. \cite{mo78},
\be
\alpha^{-1} = \left\langle {\Gamma^K_{F}\over \sum_f \Gamma^K_{f}} 
\right\rangle \left/ \left\langle {\Gamma^K_{cap}\over \sum_f \Gamma^K_{f}} 
\right\rangle\right. .
\label{WFC}
\ee
The results are shown as the dashed line in Fig. 1.  For $\Gamma^K_F =
0.105$ eV, Eq. (\ref{WFC}) gives a WFC factor of 0.43, close to that
of the full $S$-matrix treatment. However, to explain the 
observed $\alpha^{-1}$,
we have to go to much larger fission width, $\Gamma^K_F \approx 0.46$ eV,
as may be seen in Table IV. At that width
the WFC factor is 0.23 in the HF/WFC treatment and 0.16 in the
full $S$-matrix treatment.  Increasing $\Gamma^K_F$ further
does not raise the $S$-matrix value significantly.

We attribute the additional suppression in the $S$-matrix treatment to
the constraint on statistical decay rates $W_f$ imposed by 
the Bohr-Wheeler
formula \cite{bo39}
\be
W_f = \Gamma_f = {1\over 2 \pi} D T,
\ee
where $T$ is the transmission coefficient of the channel.
General considerations of detailed balance require 
$T \le 1$. The
nominal fission width in the $K$-matrix reduced width is close to the
bound, so
it is not unexpected that there is a further suppression in the $S$-matrix.

{\it Conclusion.}  We have demonstrated that branching ratios can be a
useful observable
in the study of fission dynamics near threshold.  Namely, effects not
included in the Hauser-Theory can severely constrain the number of exit 
channels.  In the example presented
here, the energy of the fissioning nucleus is above the fission barrier.
It might be of interest to apply the analysis to below-barrier fission
as well {\cite{bo13}}.
There one sees sharp peaks in the fission cross section, ascribed to
individual states along the fission path.  These states act as fission
channels with $N^{ch}_F = 1$ in the $K$-matrix modeling.

We confirmed the generality of our conclusion by exploring a variety of
channel number and transmission combinations.  Several examples are
provided in Supplemental Material.  The repository also contains
the main code implementing Eq. (1-4) and the script to compute
the branching ratio and its uncertainty.

{\it Acknowledgments.}  
The authors thank Y. Alhassid 
for discussing connections to quantum transport
theory and D. Brown for providing archived experimental data.  We also
thank W.~Nazarewicz, R. Capote, and H.A. Weidenm\"uller for very helpful
comments on an earlier version of this manuscript.


\begin{thebibliography}{99}
\bibitem{statistical} There is also an angular-momentum statistical factor
in Eq. (3) that we ignore here.  See also footnote \cite{Js}.
\bibitem{Js} There are in fact
two independent entrance channels distinguished by total angular momentum and the
actual resonance spacing is $D\approx 0.45$ \cite{ripl}. The spacing within each channel
is about twice $D_0$ \cite{mo78}.  We assume they behave similarly.
\bibitem{D} $D$ in this table includes the two independent entrance
channels. The input value for the individual entrance channels calculated 
here is about twice as large.
\bibitem{footnote} The uncertainty of the alpha ratio is much larger at higher
energies due to inconsistencies in the reported cross sections.  See
Ref. \cite{ta11}.
\bibitem{perez} The errors are estimated to be 7\% based on the data of
\cite{pe74}.
\end{thebibliography}
\end{document}